\documentclass[fleqn,usenatbib]{mnras}

\usepackage{newtxtext,newtxmath}

\usepackage[utf8]{inputenc}
\usepackage[swedish, english]{babel}
\usepackage[T1]{fontenc}
\usepackage{ae,aecompl}
\usepackage{physics}
\usepackage[dvipsnames]{xcolor}

\usepackage{graphicx}	
\usepackage{amsmath}	
\usepackage{amssymb}	

\newcommand{\Hbeta}{\mathrm{H} \beta}

\newcommand{\Msun}{\mathrm{M_{\odot}}}

\title[Balmer breaks in simulated galaxies]{Balmer breaks in simulated galaxies at z>6}

\author[Christian Binggeli et al.]   
{Christian Binggeli$^1$ \thanks{Christian.binggeli@physics.uu.se},
Erik Zackrisson$^1$,
Xiangcheng Ma$^2$,
Akio K. Inoue$^{3,4,5}$,
\newauthor
Anton Vikaeus$^1$,
Takuya Hashimoto$^{6,3,7}$,
Ken Mawatari$^{3,8}$,
Ikkoh Shimizu$^{7,9}$,
\newauthor
Daniel Ceverino$^{10,11,12}$
\\
$^1$Observational Astrophysics, Department of Physics and Astronomy, Uppsala University, Box 516, SE-751 20 Uppsala, Sweden\\
$^2$Department of Astronomy, 501 Campbell Hall 3411, University of California, Berkeley, CA 94720-3411, USA \\
$^3$Department of Environmental Science and Technology, Faculty of Design Technology, Osaka Sangyo University, \\ 3-1-1, Nagaito, Daito, Osaka 574-8530, Japan\\
$^4$Department of Physics, Faculty of Science and Engineering, Waseda University, 3-4-1 Okubo, Shinjuku, Tokyo 169-8555, Japan\\
$^5$Research Institute for Science and Engineering, Faculty of Science and Engineering, Waseda University, 3-4-1 Okubo, Shinjuku \\ Tokyo 169-8555, Japan\\
$^6$Faculty of Science and Engineering, Waseda University, 3-4-1 Okubo, Shinjuku, Tokyo 169-8555, Japan \\
$^7$National Astronomical Observatory of Japan, 2-21-1 Osawa, Mitaka, Tokyo 181-8588, Japan \\
$^8$Institute for Cosmic Ray Research, The University of Tokyo, Kashiwa, Chiba 277-8582, Japan  \\
$^9$Theoretical Astrophysics, Department of Earth \& Space Science, Osaka University, 1-1 Machikaneyama, Toyonaka, \\ Osaka 560-0043, Japan \\
$^{10}$Cosmic Dawn Center (DAWN)\\
$^{11}$Niels Bohr Institute, University of Copenhagen, Lyngbyvej 2, 2100, Copenhagen $\mbox{\normalfont\O}$, Denmark\\
$^{12}$Universit\"{a}t Heidelberg, Zentrum f\"{u}r Astronomie, Institut f\"{u}r Theoretische Astrophysik, Albert-Ueberle-Str. 2 \\ 69120 Heidelberg, Germany\\
}

\date{Accepted 2019 August 23. Received 2019 August 23; in original form 2019 May 3.}

\pubyear{2019}

\begin{document}
\def\CB{\textcolor{red}}

\label{firstpage}
\pagerange{\pageref{firstpage}--\pageref{lastpage}}
\maketitle

\begin{abstract}
Photometric observations of the spectroscopically confirmed $z\approx 9.1$ galaxy MACS1149-JD1 have indicated the presence of a prominent Balmer break in its spectral energy distribution, which may be interpreted as due to very large fluctuations in its past star formation activity. In this paper, we investigate to what extent contemporary simulations of high-redshift galaxies produce star formation rate variations sufficiently large to reproduce the observed Balmer break of MACS1149-JD1. We find that several independent galaxy simulations are unable to account for Balmer breaks of the inferred size, suggesting that MACS1149-JD1 either must be a very rare type of object or that our simulations are missing some key ingredient. We present predictions of spectroscopic Balmer break strength distributions for $z\approx 7$--9 galaxies that may be tested through observations with the upcoming \textit{James Webb Space Telescope} and also discuss the impact that various assumptions on dust reddening, Lyman continuum leakage and deviations from a standard stellar initial mass function would have on the results.
\end{abstract}




\section{Introduction}
\label{sec:intro}
Characterizing galaxy properties at the very highest redshifts is important for understanding how galaxy formation and evolution proceeded across cosmic time. Much of our current insight about star formation in galaxies at the high-redshift universe, however, comes from numerical simulations. While these are powerful tools, currently, the lack of high-quality spectra at the highest redshifts makes it difficult to calibrate and verify predictions made by simulations. In the near future, the upcoming \textit{James Webb Space Telescope} (\textit{JWST}) may be able to shed light on many different properties of the high-redshift galaxy population that are beyond the reach of current facilities. At the time of writing, there are only two objects that have been spectroscopically confirmed at $z\gtrsim 9$; GN-z11 \citep{oesch_remarkably_2016} and MACS1149-JD1 \citep[hereafter JD1; ][]{zheng_magnified_2012,huang_spitzer_2016,kawamata_precise_2016,zheng_young_2017, hoag_hst_2018, hashimoto_onset_2018} While the redshift of GN-z11 ($z=11.09$) was determined using the position of a continuum break in its spectrum (seen in \textit{Hubble Space Telescope} (\textit{HST}) grism data), JD1 represents the most distant object for which emission lines have been detected. Using observations of the [OIII] 88-$\mathrm{\mu m}$ and Lyman-alpha emission lines performed with the Atacama Large Millimeter/Submillimeter Array (ALMA) and X-shooter at the Very Large Telescope (VLT), \citet{hashimoto_onset_2018} were able to constrain the redshift of JD1 to $z = 9.11$. An interesting feature of this object is that it exhibits a red color in the \textit{Spitzer}/IRAC 3.6 and 4.5-$\mathrm{\mu m}$ channels (channel 1 and 2). While \citet{zheng_magnified_2012} suggested that the red color could be a sign of a prominent Balmer break around 4000 Å rest-frame, other studies of objects with similar \textit{Spitzer}/IRAC colors have suggested that the red color in these objects is likely due to strong [OIII] and $\Hbeta$ emission \citep[e.g.][]{roberts-borsani_z_2016,stark_ly_2017}. In the case of JD1, earlier studies have not been able to determine the exact source of the red color due to the uncertainty in the redshift \citep[e.g.][]{hoag_hst_2018}. However, with the spectroscopic redshift determination by \citet{hashimoto_onset_2018}, it became possible to rule out that the excess in the 4.5-$\mathrm{\mu m}$ channel arises from a strong [OIII]~$\lambda$5007 emission line, leading to the conclusion that the object exhibits a strong Balmer break.

With a detailed analysis combining observations over a wide wavelength range, \citet{hashimoto_onset_2018} found that the spectral energy distribution (SED) of JD1 is consistent with the object having a star formation history (SFH) with two episodes of high star formation and an intermediate phase (with a duration of $\sim 100 \text{ to } 200$ Myr) of low/no star formation activity. An alternative interpretation is that the Balmer break of JD1 is primarily due to selective dust attenuation, in the sense that the light from young stars is far more heavily extinguished than the light from old stars \citep{katz_probing_2019}. There are other examples of galaxies that seem to exhibit evolved features such as strong Balmer breaks at lower redshifts \citep[$z\gtrsim 5$][]{mawatari_possible_2016}.
Furthermore, there are galaxies at $z\sim 9$ that currently lack spectroscopic redshift confirmations which exhibit a similar \textit{Spitzer}/IRAC color to that seen in JD1 \citep{oesch_most_2014}. Several different cosmological simulation suites predict that galaxies in the early universe should experience periodic, or so called `bursty', star formation intermitted by periods of low star formation activity \citep[e.g.][]{kimm_towards_2015,trebitsch_fluctuating_2017, hopkins_fire-2_2018, ma_simulating_2018, ceverino_firstlight_2018, ma_dust_2019}. It is, however, unclear whether the variations in SFR seen in simulations are large enough to produce Balmer breaks as strong as the one seen in JD1. Future \textit{JWST} observations should be able to constrain the Balmer break strengths of a large number of high-redshift galaxies. By comparing these to the Balmer breaks exhibited by simulated galaxies, one should hence be able to determine if simulations accurately reproduce galaxies in the high-redshift universe, and calibrate the simulations in such cases that they do not. In this study, we utilize simulated galaxies from the \citet{shimizu_nebular_2016} and FIRE-2 simulations \citep{ma_simulating_2018, ma_dust_2019} in order to make predictions of Balmer break strengths for a large sample of high-redshift galaxies. We also compare our predictions to those obtained with the FirstLight simulation \citep{ceverino_firstlight_2018}. 

This paper is organized as follows. We briefly describe the simulated sample of galaxies in section~\ref{sec:simulations}. In section~\ref{sec:Synthetic_SEDs} we explain our method for generating synthetic spectra for the simulated galaxies, and discuss assumptions in our model. We introduce a simple diagnostic for the Balmer break especially aimed at the observing capabilities of the upcoming \textit{JWST} in section~\ref{sec:method_balmer_break}. Our results and the effects of assumptions regarding dust, escape of ionizing photons, simulation choice and the initial mass function (IMF) are presented in section~\ref{sec:predicted_Balmer_breaks} and section~\ref{sec:dust_n_SED}. Section~\ref{sec:JD1} focuses mainly on the size of the Balmer breaks in the simulated galaxies as measured by \textit{Spitzer}/IRAC, and how these compare to the Balmer break observed in JD1. Finally, in section~\ref{sec:Disc_BBJWST}, we discuss implications of the results obtained in this study and discuss our results in the context of other recent studies on the topic.

\section{Model}
\label{sec:model}
In this section, we briefly describe the cosmological simulations, how the simulated galaxies are selected and how we generate synthetic SEDs for the galaxies. 
\subsection{Simulated galaxies}
\label{sec:simulations}

In this study, we utilize simulated galaxies from a total of three independent cosmological simulation suites in order to get a sample of realistic epoch-of-reionization (EoR) galaxies. The first two sets of galaxies come from the FIRE-2 (Feedback In Realistic Environments) simulations, described in \citet{ma_simulating_2018, ma_dust_2019} and the simulations described in \citet[][hereafter S16]{shimizu_nebular_2016}. The S16 simulation is based on the smoothed particle hydrodynamics code \textsc{gadget-3}, which is an updated version of the \textsc{gadget-2} code \citep{springel_cosmological_2005}. For the set of galaxies used in this study, the simulation is run over a volume of $50^3 \  (h^{-1}\mathrm{Mpc})^3$ with a mass resolution of $\approx 6.6 \ \times 10^6 \ \Msun$ for dark-matter particles, while the initial mass for star-particles is $\approx 3 \times 10^5 \ \Msun$. The FIRE-2 sample used here comes from a suite of cosmological zoom-in simulations that are run using the \textsc{gizmo} code \citep{hopkins_new_2015} in a $120^3 \ (h^{-1}\mathrm{Mpc})^3$ box, with a dark-matter particle resolution of $\approx 4 \times 10^4 \ \Msun$ and an initial mass for star-particles of $\approx 7000 \ \Msun$. In the FIRE-2 simulations, galaxies are selected from high-resolution zoom-in regions centered around massive halos ($M_{halo} \sim 10^{11}-10^{12} \Msun$) within the simulation volume. For the set of simulations used here, 6 zoom-in regions are selected and run with higher resolution to $z=7$ and an additional 6 regions are selected from an independent volume and run to $z=9$ \citep[see][]{ma_dust_2019}. While the total volume of the FIRE-2 simulations is larger than the volume in the S16 simulation, the zoom-in regions limit the number of objects extracted, leading to significantly fewer objects with unique SFHs at the relevant redshifts. Thus, the S16 simulation has the advantage of providing us with a large number of galaxies with individual star formation histories, while the FIRE-2 simulations provide us with fewer, but more highly resolved objects. Furthermore, the galaxies within the FIRE-2 simulations exhibit significantly bursty SFHs over time \citep[see][]{ma_simulating_2018}. We are thus more likely to catch galaxies in, or around a period of low star formation activity, where Balmer breaks are likely to be a more prominent feature of the SED. SFHs and stellar metallicity distributions are extracted from the simulated galaxies with stellar masses $M_{\star} > 10^8 \ \Msun$ at $z\sim 7-9$. For the S16 simulations, we extract this information in snapshots taken at $z=9,8 \ \text{and} \ 7$, while for the FIRE-2 simulations, we bin several snapshots into $\Delta z \sim 1$ bins around $z=9,8 \ \text{and} \ 7$ in order to increase the total number of galaxies. Our criteria yield a total of 182, 513 and 1277 galaxies for redshifts $z=9,8 \ \text{and} \ 7$ respectively from the S16 simulations, and 150, 152 and 187 galaxies at the corresponding redshifts for the FIRE-2 simulations. In the case of the FIRE-2 simulation, the individual snapshots (before we re-bin in redshift) contain between 13 and 28 objects with unique SFHs. Due to the zoom-in method used in the FIRE-2 simulations, and the fact that the number of zoom-in regions at $z=9$ is twice the number of zoom-in regions at $z=7$, the number of galaxies extracted at increasing redshifts does not drop off as rapidly as in the S16 simulations.

As a third set, we utilize galaxies from the FirstLight project \citep{ceverino_introducing_2017, ceverino_firstlight_2018, ceverino_firstlight_2019}. The FirstLight galaxy sample comes from a homogeneous, mass-complete suite of cosmological zoom-in simulations that are run using the \textsc{art} code \citep[e.g.][]{kravtsov_adaptive_1997,ceverino_role_2009,ceverino_radiative_2014}, and the sample used here is extracted from simulations volumes of $10^3 \  (h^{-1} \mathrm{Mpc})^3$ and $20^3 \ (h^{-1} \mathrm{Mpc})^3$. In this case, haloes with masses $\sim 10^9 \text{ to } 10^{11} \Msun$ are selected from a low-resolution simulation at $z=5$ and simulated at higher resolution. The resolution in this simulation is $10^4 \ \Msun$ for dark-matter particles while the minimum initial mass for star-particles in the simulation is $100 \ \Msun$ \citep[see][]{ceverino_firstlight_2019}. A fundamental difference between the FirstLight sample and the FIRE-2/S16 sample is that we utilize pre-generated stellar and nebular SEDs that are publicly available\footnote{\url{http://www.ita.uni-heidelberg.de/~ceverino/FirstLight/}}, meaning that the SEDs are not generated via the method described in section~\ref{sec:Synthetic_SEDs}. Before calculating Balmer break strengths, we perform the same mass cut as for the other two simulation sets, i.e. require $M_{\star} > 10^8 \ \Msun$ and combine SEDs of galaxies from snapshots $8.5<z<9.5$ in order to increase our sample size. This leaves us with a total of $83$ SEDs at $z=9$. As in the case with the FIRE-2 simulations, the FirstLight simulation gives us a smaller sample than the S16 simulation, but has the advantage of a significantly higher resolution. Similar to the FIRE-2 simulations, these also exhibit bursty SFHs, and are thus also likely to produce galaxies with prominent Balmer breaks. The FirstLight SEDs are generated using BPASS v.2.1 binary evolution model spectra \citep{eldridge_binary_2017}, and are generated without taking any effects of dust-reddening into account \citep[see][]{ceverino_firstlight_2019}. As this set of SEDs is generated without considering dust-reddening and different assumptions regarding stellar evolution, we compare this set to a set of dust-free FIRE-2 galaxies at $z\sim 9$ (see section~\ref{sec:dust_n_SED-simulation_choice}) to test whether they are consistent. Since the simulations suites are based on different assumptions and numerical implementations, the use of several independent simulations ensure that our conclusions are generic, or at least not limited to any single batch of models.

\subsection{Synthetic spectra of simulated galaxies}
\label{sec:Synthetic_SEDs}
In order to generate synthetic spectra for the simulated galaxies, we utilize a grid of single-age stellar population (SSP) spectra from Yggdrasil \citep{zackrisson_spectral_2011} spanning over a wide range in metallicity $\mathrm{(Z)}$ and age $\mathrm{(t)}$. In this study, this grid consists of \textsc{starburst99} Padova-AGB models \citep{leitherer_starburst99:_1999,vazquez_optimization_2005} for metallicities $Z=0.0004 \textrm{ -- } 0.040$ scaled to a \citet{kroupa_variation_2001} universal IMF. For each such single stellar population, the associated nebular emission is calculated using the \textsc{cloudy} photo-ionization code \citep{ferland_2013_2013} while assuming a spherical nebula with constant density ($n(\mathrm{H}) = 100 \ \mathrm{cm}^{-3}$) and that the nebular metallicity is equal to the stellar metallicity \citep[for further details, see][]{zackrisson_spectral_2011}. At this stage, we can re-scale the nebular contribution to the overall SED in order to account for nonzero escape fraction of ionizing photons. While we explore effects of varying the escape fraction, we use an escape fraction of zero as the default option in this study. In order to find a spectrum for each star-particle within the simulated galaxies, we interpolate the grid of SSP spectra in $\mathrm{log(t)}$ and $\mathrm{log(Z)}$. In order to obtain the total spectrum of the galaxy, we then simply sum over all star-particles belonging to each individual galaxy. 

Since dust may have a strong effect on the emergent SED and strength of the Balmer break, we also consider dust-reddening effects on the synthetic spectra. In the case of the S16 simulations, a galaxy-wide dust extinction at 1500 Å is provided for each object from the simulation. This prediction is then used to scale a \citet{pei_interstellar_1992} Small Magellanic Cloud (SMC) extinction curve in order to redden the spectrum. 
For one set of the FIRE-2 galaxies, we apply the same SMC curve while assuming a fixed extinction in the V band of $A_{\mathrm{V}}=0.5$ magnitudes for all galaxies. Note that this dust recipe is quite extreme, and produces galaxies that have significantly redder ultraviolet (UV) spectral slopes than those observed at high redshifts \citep[e.g.][]{bouwens_uv-continuum_2014}. This crude dust recipe does, however, provide us with a way to test the effect of a galaxy-wide dust extinction on the distribution of Balmer break strengths. In both of these cases, dust extinction is occurring in a so-called dust-screen around the galaxy. While this recipe if often used to account for reddening, more realistic dust recipes should take the distribution of dust into account. For this purpose, we also utilize a version of the FIRE-2 galaxies that have been post-processed in the \textsc{skirt} dust radiative transfer code \citep{camps_skirt:_2015}. The \textsc{skirt} post processing code is run while assuming an SMC-type dust grain distribution \citep{weingartner_dust_2001}, a dust-to-metal ratio of $0.4$ and that no dust is present in gas hotter than $10^6$ K. In this case, we show the Balmer break strengths predicted for the emergent spectrum along a randomly selected line of sight. This version of the FIRE-2 simulations with \textsc{skirt} post-processing are presented in \citet{ma_dust_2019}. We also utilize a version of the FIRE-2 galaxies that have undergone post processing in \textsc{skirt} while assuming a higher dust-to-metal ratio of $0.8$. The motivation for this is that these provide a better match to observed UV luminosity functions at $z>6$. Increasing the dust-to-metal ratio can also be seen as a proxy for boosting dust opacity and provides a way to test the effect of an increased dust opacity on observed Balmer breaks \citep[see][]{ma_dust_2019}. In the post-processing scheme, a different (but very similar) grid of \textsc{starburst99} Padova-AGB SSP spectra from \textsc{sunrise} \citep{jonsson_high-resolution_2010} is utilized. This grid only includes nebular emission (continuum and lines) from hydrogen. In order to compare the resulting Balmer breaks to observations of JD1, we calculate synthetic fluxes in \textit{Spitzer}/IRAC 3.6 and 4.5 $\mathrm{\mu m}$ channels for the simulated galaxies at $z \sim 9$. This is done by integrating the synthetic SEDs over the IRAC filter curves assuming all simulated galaxies are at the redshift of JD1 ($z=9.11$). In the \textsc{skirt} post processing procedure, the SEDs are interpolated to a lower wavelength resolution for computational purposes. An effect of this coarser grid of wavelength points is that contribution from emission/absorption lines is not accounted for when the flux in the IRAC filters is calculated. However, since the spectra in this case only include nebular emission lines and continuum from hydrogen, this does not have a significant impact on the predicted IRAC fluxes. For the rest of the study, unless it is otherwise specified, the default option for the \textsc{skirt} post-processed galaxies is a dust-to-metal ratio of 0.4.

\subsection{The Balmer break}
\label{sec:method_balmer_break}

\begin{figure}
    \begin{center}
    \includegraphics[width=0.48\textwidth]{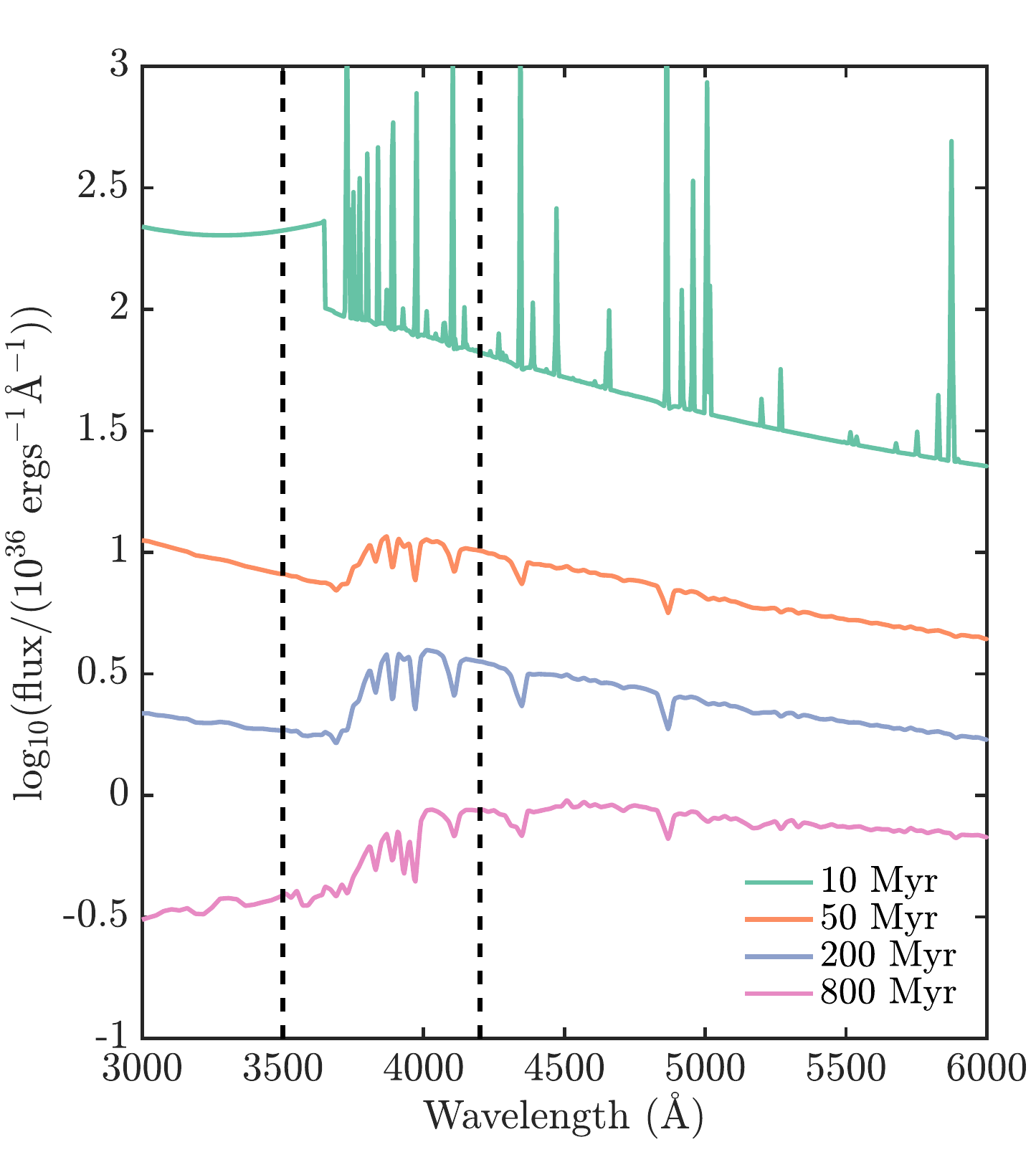}
    \end{center}
    \caption{Yggdrasil SSP spectra generated with Padova-AGB models for 10 (cyan, top), 50 (orange, second from top), 200 (blue, second from bottom) and 800 (pink, bottom) Myr stellar populations with metallicity $\mathrm{Z}=0.02$. The spectra have been shifted in the flux direction for clarity. The two dashed black lines mark wavelengths that are suitable to determine the Balmer break strength using \textit{JWST}/NIRSpec.}
    \label{fig:Padova_SSPs}
\end{figure}

One of the goals of the upcoming \textit{JWST} is to study the galaxy population at high redshifts. Using the NIRSpec instrument, we should be able to get accurate measurements of Balmer break strengths for galaxies brighter than $\mathrm{m_{AB}(1500\text{Å})}=27 \text{ mag}$ within $\approx 10$ hours of exposure (see section~\ref{sec:Disc_BBJWST}). In order to discuss Balmer break distributions in a way that is relevant for the \textit{JWST}, we calculate Balmer breaks defined as the flux (in units of $F_{\nu}$) at 4200 Å over the flux at 3500 Å:

\begin{equation}\label{eq:B_JWST_Def}
    B_{\mathrm{4200/3500}}=\frac{F_{\nu}(4200 \text{Å})}{F_{\nu}(3500 \text{Å})}
\end{equation}

Fig~\ref{fig:Padova_SSPs} shows Yggdrasil (Padova-AGB) SSPs at different ages of the stellar population, with these wavelengths marked in dashed lines. Continuum fluxes at these wavelengths should be observable with the \textit{JWST}/NIRSpec with the lowest resolution setting ($R \sim 100$), thanks to the high sensitivity of this mode. Even with this low resolution, fluxes at 3500 Å and 4200 Å should be observable at a wide redshift range without the risk of absorption/emission lines blending into the relevant spectral bins. As mentioned in the above section, for comparison with observations of JD1, we also calculate Balmer breaks as observed by \textit{Spitzer}/IRAC by integrating our synthetic SEDs over the IRAC channel 1 and 2 transmission curves. 

\section{Predicted Balmer break strength distributions}
\label{sec:predicted_Balmer_breaks}

\begin{figure}
\begin{center}
\includegraphics[width=0.48\textwidth]{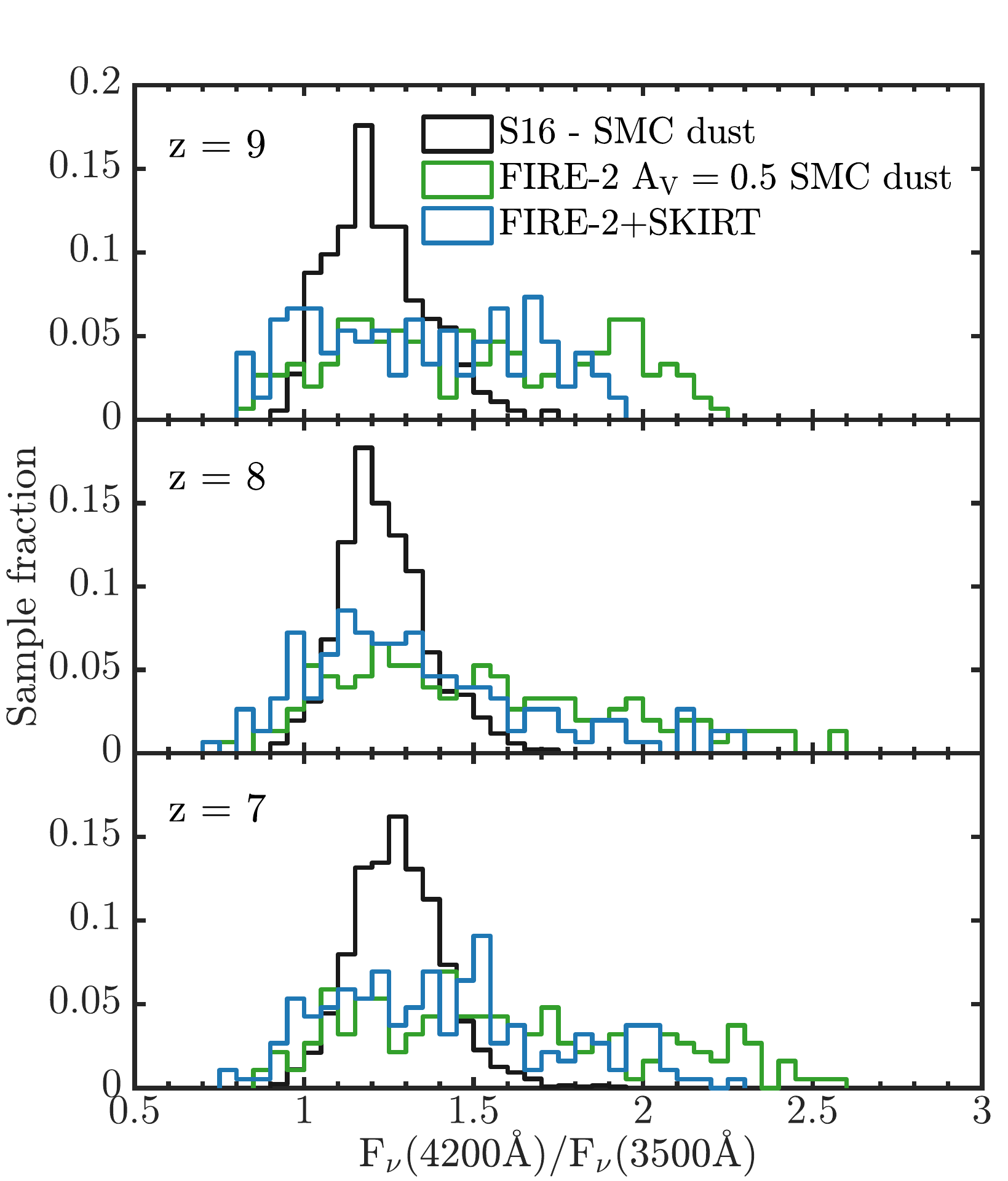}
\end{center}
\caption{Distributions of Balmer break strength ($ F_{\nu}(4200 \text{Å})/F_{\nu}(3500 \text{Å}) $, $ B_{\mathrm{4200/3500}}$ in the text) at $z=9$ (top), $z=8$ (middle) and $z=7$ (bottom) of simulated galaxies with stellar masses $M_{\star} \geq 10^8 \ \Msun$. The green distribution shows the FIRE-2 galaxies with Yggdrasil Padova-AGB SSP grid and SMC dust reddening with a fixed extinction of $0.5$ mag in the V-band. The blue distribution shows the FIRE-2 galaxies after post-processing with \textsc{skirt} and the black distribution shows the S16 galaxies with SMC dust extinction. The total number of galaxies at each redshift ($z=9,8,7$) is 182, 513 and 1277 for the S16 simulation, and 150, 152 and 187 for the FIRE-2 simulation.}
\label{fig:Fnu4200_3500_z_9_8_7}
\end{figure}

Fig~\ref{fig:Fnu4200_3500_z_9_8_7} shows the Balmer break strength distributions as defined in equation~\ref{eq:B_JWST_Def} at redshifts 9, 8 and 7 for dusty galaxies with masses $M_{\star}>10^8 \ \Msun$. Properties of the Balmer break strength distributions are shown in table~\ref{tab:Result_z987}. Overall, the larger variations in the past star formation activity in the FIRE-2 galaxies produces distributions of Balmer breaks that are significantly wider than those seen in the S16 simulation. The reason for this is that the larger variety in SFHs in the FIRE-2 simulation leads to a larger range in galaxies with low versus high star formation rate. This simply means that we are more likely to find galaxies that have experienced a rapid drop or increase in star formation rate, which ultimately affects the Balmer break strength. While this is true for either of the dust recipes used for the FIRE-2 galaxies, the extreme test case of a fixed extinction leads to significantly larger Balmer breaks for a subset of the galaxies. Furthermore, both of the FIRE-2 Balmer break distributions (see fig~\ref{fig:Fnu4200_3500_z_9_8_7} and table~\ref{tab:Result_z987}) have higher mean values than the S16 distribution. For either of the simulations, the mean and median of the distributions is not affected strongly with decreasing redshift. However, for the FIRE-2 simulations, the fraction of objects with the largest Balmer breaks ($B_{\mathrm{4200/3500}}>2$) increases slightly with decreasing redshift. For the S16 simulation, no galaxies exhibit Balmer breaks stronger than $B_{\mathrm{4200/3500}}=2$ at any redshift. In principle, it should be fairly straightforward to compare the distributions in fig~\ref{fig:Fnu4200_3500_z_9_8_7} to \textit{JWST}/NIRSpec observations of galaxies at $z>6$ and thus get a way to test if simulated and observed Balmer break strengths match.

\begin{table}
    \centering
    \caption{Mean, median, 5th and 95th percentile of the Balmer break ($ B_{\mathrm{4200/3500}}$) distributions and fraction of galaxies with Balmer breaks stronger than  $ B_{\mathrm{4200/3500}}=2$ ($f_{>2}$) at redshifts $9, \ 8 \text{ and } 7$ for the different models.}
\begin{tabular}{c c c c c c}
\hline
\multicolumn{6}{c}{S16 - SMC dust}\\
redshift & mean & median & 5 pctl. & 95 pctl. & $f_{>2}$ \\
\hline
9 & 1.22 & 1.20 & 1.01 & 1.47 & 0.00 \\
8 & 1.24 & 1.22 & 1.03 & 1.49 & 0.00 \\
7 & 1.28 & 1.27 & 1.08 & 1.51 & 0.00 \\
\multicolumn{6}{c}{ }\\
\multicolumn{6}{c}{FIRE-2 $A_{\mathrm{V}}=0.5$ SMC dust}\\
\hline
9 & 1.51 & 1.49 & 0.93 & 2.10 & 0.11 \\
8 & 1.52 & 1.47 & 0.98 & 2.30 & 0.14 \\
7 & 1.59 & 1.52 & 1.01 & 2.31 & 0.21 \\
\multicolumn{6}{c}{ }\\
\multicolumn{6}{c}{FIRE-2+SKIRT}\\
\hline
9 & 1.34 & 1.32 & 0.87 & 1.83 & 0.00 \\
8 & 1.34 & 1.27 & 0.93 & 2.09 & 0.06 \\
7 & 1.41 & 1.39 & 0.95 & 2.03 & 0.06 \\
\hline
\end{tabular}
    \label{tab:Result_z987}
\end{table}

\begin{figure*}
    \begin{center}
    \includegraphics[width=0.95\textwidth]{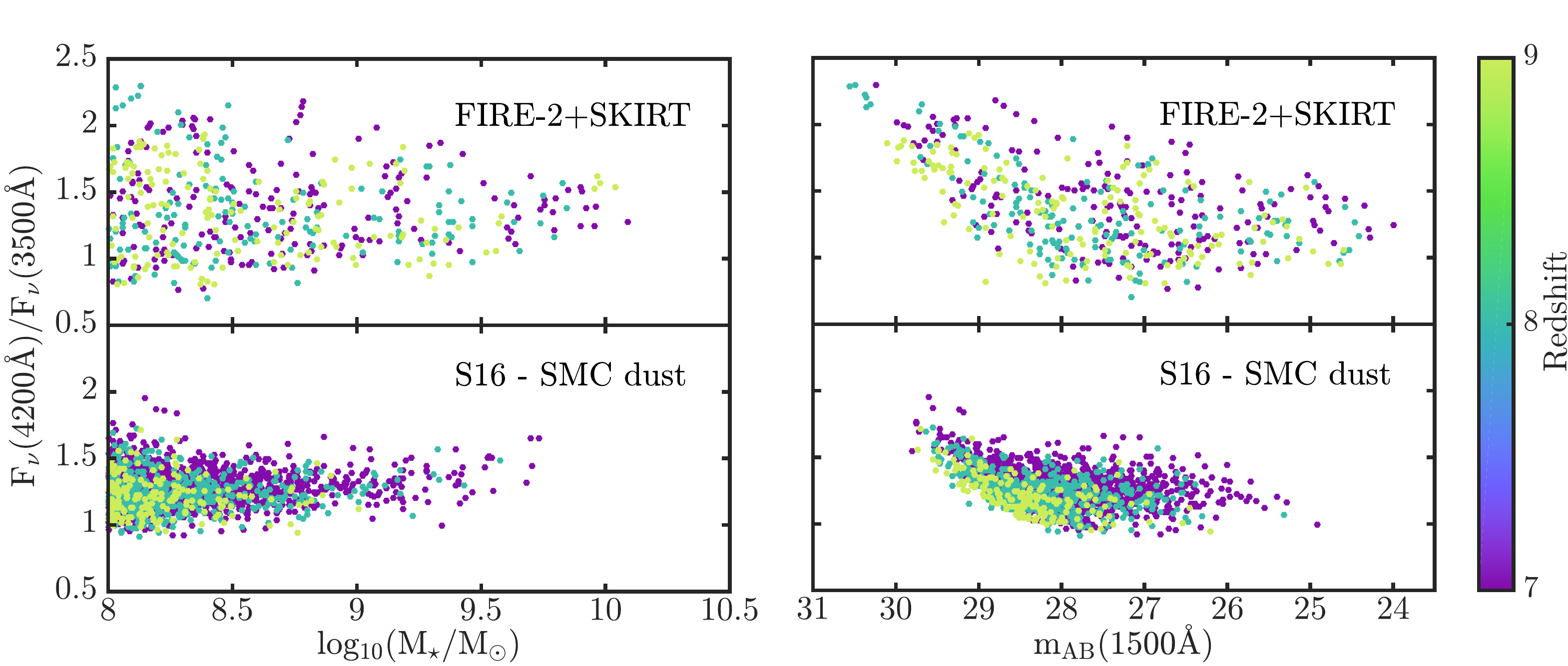}
    \end{center}
    \caption{Total stellar mass (left panel) and apparent UV magnitude ($\mathrm{m_{AB}(1500\text{Å})}$; right panel) vs Balmer break strength ($ F_{\nu}(4200 \text{Å})/F_{\nu}(3500 \text{Å}) $, $ B_{\mathrm{4200/3500}}$ in the text) vs redshift (color axis) for the FIRE-2 simulations that have undergone post-processing in \textsc{skirt} (top) and the S16 galaxies (bottom). The total number of galaxies is 182, 513 and 1277 for the S16 simulation, and 150, 152 and 187 for the FIRE-2 simulation at redshift 9, 8 and 7 respectively.}
    \label{fig:Mass_UVmag_Fnu4500_3500}
\end{figure*}

Fig~\ref{fig:Mass_UVmag_Fnu4500_3500} shows the Balmer break strengths versus stellar mass and observed UV magnitude at redshifts 9, 8 and 7. In this figure, we only show the S16 simulations and the FIRE-2 galaxies that have undergone post processing in \textsc{skirt}. The predicted distributions of Balmer break strengths in this figure widen with decreasing mass, such that the largest variety of Balmer break strengths is observed for the lowest mass galaxies. While this can be explained with more bursty star formation in the low-mass galaxies, this could also be an effect of the limited sample size at higher masses. While the Balmer break strength is not strongly affected by assuming a higher dust-to-metal ratio of 0.8, the observed UV magnitudes can change as much $\sim 0.5$ (see appendix~\ref{sec:appendix}).

\subsection{Modelling assumptions \&  Caveats}
\label{sec:dust_n_SED}
In the following sections, we investigate effects of the assumptions made in our modelling. We discuss effects of dust reddening, non-zero escape fractions of ionizing photons, stellar evolutionary modelling and simulation choice.

\subsubsection{Dust reddening effects}
\label{sec:dust_n_SED-Dust}
\begin{figure}
\begin{center}
\includegraphics[width=0.48\textwidth]{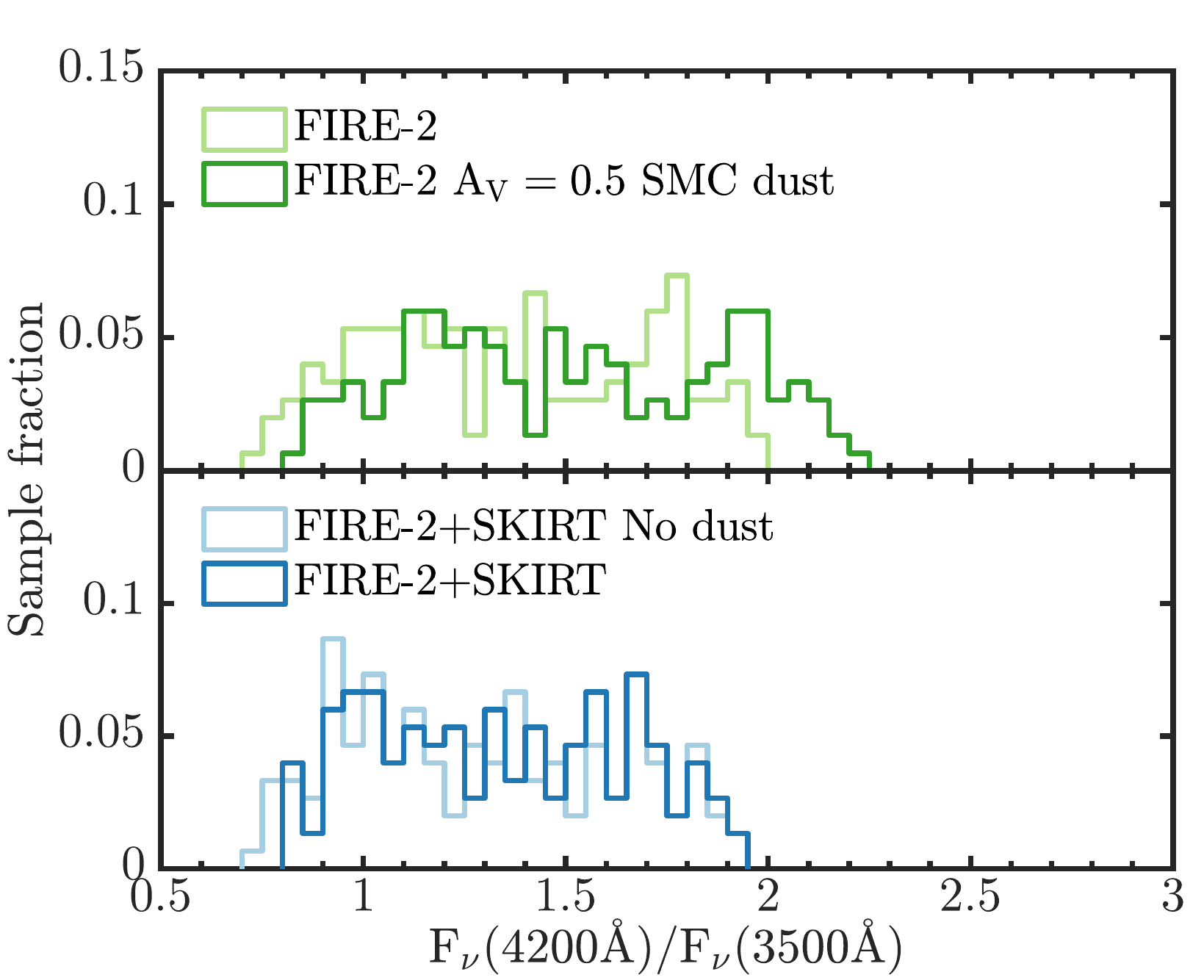}
\end{center}
\caption{Distribution of Balmer break strength ($ F_{\nu}(4200 \text{Å})/F_{\nu}(3500 \text{Å}) $, $ B_{\mathrm{4200/3500}}$ in the text) for different dust treatments at $z = 9$. The top panel shows The FIRE-2 galaxies where spectra have been generated using the Yggdrasil Padova-AGB SSP grid without dust (light green) and with SMC dust reddening given a fixed extinction of $0.5$ mag in the V-band (dark green). The bottom panel shows the FIRE-2 galaxies after post processing in \textsc{skirt}. The distributions show the emergent SED without dust (light blue) and with dust effects taken into account (dark blue). Each distribution contains 150 objects.}
\label{fig:Fnu_Dusteffect}
\end{figure}

The effect of dust reddening on the simulated $B_{\mathrm{4200/3500}}$ distributions for the FIRE-2 galaxies at $z=9$ are shown in fig~\ref{fig:Fnu_Dusteffect}. While the \textsc{skirt} post-processing leads to age-dependent differential obscuration in the galaxies (meaning that younger stars generally experience larger dust obscuration than old stars), the effect on the Balmer break strength distributions is small. While individual galaxies can experience an increase in the exhibited Balmer break of $  \Delta B_{\mathrm{4200/3500}} \sim 0.4$ in the dusty case compared to the dust-free case, the overall distribution is not strongly affected. Generally, the \textsc{skirt} post-processing leads to a larger dust extinction in massive galaxies than in the low mass galaxies. This means that the Balmer break strength is only weakly affected in a majority of galaxies, and in particular, in the galaxies with the largest dust-free Balmer breaks (i.e. the low-mass ones).

As mentioned in section~\ref{sec:predicted_Balmer_breaks}, increasing the dust-to-metal ratio to 0.8 does not significantly change this result (see appendix~\ref{sec:appendix}). Note that we only present results for one random line-of-sight for the FIRE-2 galaxies that have undergone post-processing in \textsc{skirt}. We have, however studied several different random lines-of-sight for all of the galaxies, and find no significant difference to the results presented here. Balmer break strengths for the FIRE-2 galaxies calculated using the Yggdrasil SSP grid are also shown in fig~\ref{fig:Fnu_Dusteffect}. The mean, median,  5, 95 percentile. and fraction of galaxies with $B_{\mathrm{4200/3500}} > 2$ are $1.35$, $1.34$, $0.84$, $1.88$ and $0$ for the Yggdrasil SSP case when dust effects are ignored. Corresponding values for the \textsc{skirt} post-processed case without dust are $1.30$, $1.28$, $0.81$, $1.83$ and 0. The $A_{\mathrm{V}}=0.5$ SMC recipe case assumes that dust is distributed in a screen around the stars, and that all stars experience the same amount of reddening. While invoking this dust recipe with such a large extinction can lead to significantly stronger Balmer breaks, as explained in section~\ref{sec:Synthetic_SEDs}, this type of recipe is not consistent with observed UV slopes at high redshifts.

\subsubsection{Escape fraction effects}
\label{sec:dust_n_SED-fesc}
Setting the escape fraction of ionizing photons to unity shifts the simulated $ B_{\mathrm{4200/3500}}$ distribution towards higher Balmer breaks. The average shift for the S16 galaxies at $z=9$ is about $\Delta  B_{\mathrm{4200/3500}}\approx 0.25$, with one galaxy experiencing an increase of $\Delta  B_{\mathrm{4200/3500}} \approx 0.35 $.  For the FIRE-2 galaxies with the Yggdrasil SSP grid and $A_{\mathrm{V}}=0.5$ dust recipe at $z=9$, the corresponding average shift is $\Delta  B_{\mathrm{4200/3500}}\approx 0.21$ with the largest individual shift of $\Delta  B_{\mathrm{4200/3500}}\approx 0.50$. We observe the largest difference in Balmer break as an effect of escape fraction for the galaxies that have the weakest Balmer breaks. In the case the S16 simulation, the fraction of galaxies that exhibit a Balmer break larger than $2$, remains at zero, while it increases slightly, to $0.19$ for the FIRE-2 galaxies. In the case that dust-effects are ignored, the largest individual and average increase in the Balmer break for the FIRE-2 galaxies with the Yggdrasil SSP grid is  $\Delta  B_{\mathrm{4200/3500}} \approx 0.19$ and  $\Delta  B_{\mathrm{4200/3500}} \approx 0.45$, respectively. The fraction of galaxies that exhibit $B_{\mathrm{4200/3500}}>2$ increases from $0$ to $0.01$ when the escape fraction is increased from 0 to unity for the dust-free case. The change in the exhibited Balmer break can be attributed to the change in the nebular continuum emission blueward of the Balmer break as the escape fraction changes.

\subsubsection{Simulation choice}
\label{sec:dust_n_SED-simulation_choice}
\begin{figure}
\begin{center}
\includegraphics[width=0.48\textwidth]{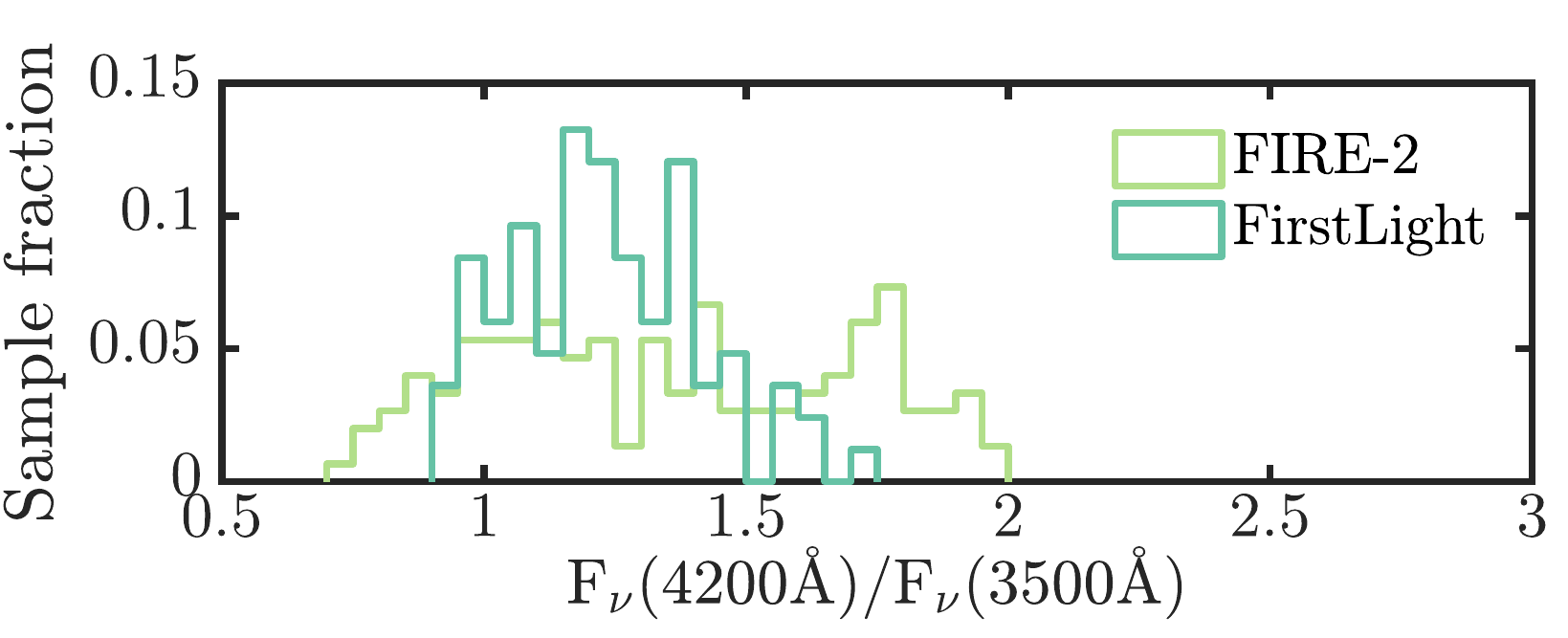}
\end{center}
\caption{Balmer break strength ($ F_{\nu}(4200 \text{Å})/F_{\nu}(3500 \text{Å}) $, $ B_{\mathrm{4200/3500}}$ in the text) distributions for dust-free simulated galaxies at $z = 9$. The figure shows 150 galaxies from the FIRE-2 simulation where spectra have been generated using the Yggdrasil Padova-AGB SSP grid (light green) and 83 galaxies from the FirstLight simulation (cyan, see section~\ref{sec:dust_n_SED}).}
\label{fig:BalmerBreak_FIRE2_FirstLight}
\end{figure}

As mentioned in section~\ref{sec:simulations}, we also utilize dust-free synthetic SEDs from the FirstLight simulation. The distribution of Balmer breaks as measured by $  B_{\mathrm{4200/3500}}$ for this simulation set is shown in figure~\ref{fig:BalmerBreak_FIRE2_FirstLight}. In the same figure, we also show $  B_{\mathrm{4200/3500}}$ for dust-free synthetic FIRE-2 SEDs generated with the Yggdrasil Padova-AGB SSP grid. While the Balmer breaks obtained using the FirstLight SEDs is centered at slightly weaker Balmer breaks, the main difference is in the width of the distributions. Considering the difference in modelling assumptions and sample sizes, conclusions drawn from the two simulations would be similar. The mean, median, 5 and 95 percentile. for the FirstLight set shown in figure~\ref{fig:BalmerBreak_FIRE2_FirstLight} is $1.23$, $1.23$, $0.96$, $1.57$, respectively, while no objects exhibit $B_{\mathrm{4200/3500}}>2$. 
As discussed in section~\ref{sec:simulations}, the FirstLight SEDs are generated using BPASS v.2.1. binary stellar evolutionary spectra which predict a higher ionizing flux compared to the \textsc{starburst99} models. This means that the Balmer break may be slightly weaker due to increased nebular continuum emission around 3500. Furthermore, due to mass transfer between binaries, BPASS binary models exhibit a decrease in red supergiant stars, and an increase in hot stripped stars. This may also lead to weaker Balmer breaks in binary models \citep{ma_simulating_2018}.

\subsection{IMF effects}
\label{sec:IMF_effects}
Concerns regarding the universality of the IMF \cite[e.g.][]{kroupa_variation_2001, conroy_stellar_2012, conroy_stellar_2017, van_dokkum_stellar_2017} and its possible effect on the Balmer break strengths in the simulated galaxies justifies investigating different IMFs. We consider the IMF as a set of power-laws for the different ranges of stellar masses;

\begin{equation} \label{eq:IMF}
\dv{N}{M} \propto M^{-\alpha}
\end{equation}

\begin{figure}
\begin{center}
\includegraphics[width=0.475\textwidth]{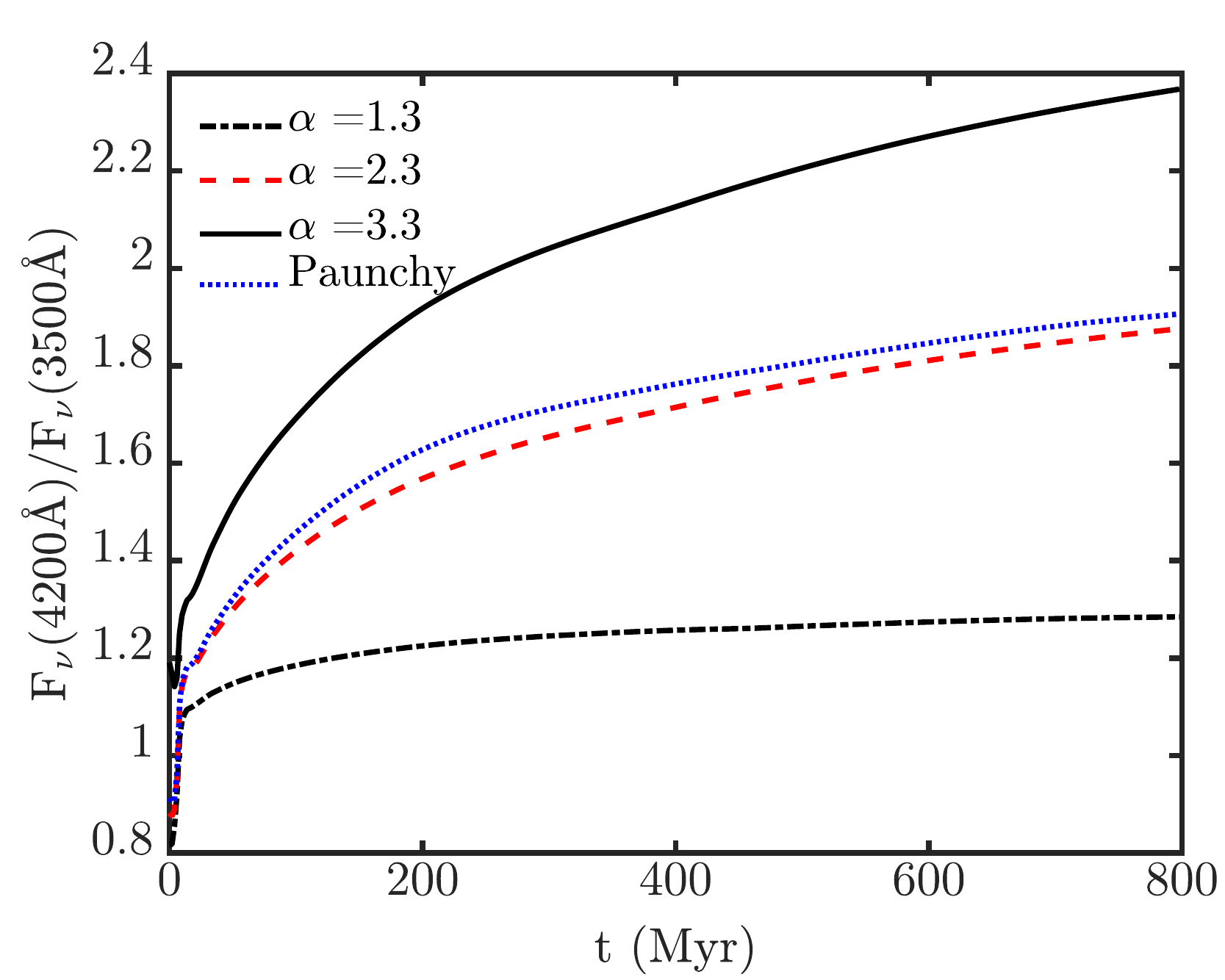}
\end{center}
\caption{Balmer break strength ($ F_{\nu}(4200 \text{Å})/F_{\nu}(3500 \text{Å}) $, $ B_{\mathrm{4200/3500}}$ in the text) as a function of age (t), calculated using SSP spectra from a \textsc{starburst99} Padova-AGB model using four different configurations of the IMF power-law (\autoref{eq:IMF}). The blue dotted line shows a `paunchy' IMF, while the other three show two-component IMFs with power-law slopes $\alpha$ set to 1.3 in the mass range $M =  \text{ 0.1 -- 0.5 }\Msun$ and $\alpha = 1.3$ (black, dash-dotted), 2.3 (red, dashed) and 3.3 (black, solid) for the mass interval $M= \text{ 0.5 -- 100 } \ \Msun$.}
\label{fig:Bbreak_IMF_PadAgb}  
\end{figure}

Where $N$ is the number of stars, $M$ is the stellar mass and $\mathrm{\alpha}$ is the power-law slope of the IMF. In order to estimate effects of the IMF on the resulting Balmer break, we utilize \textsc{starburst99} and consider a Padova-AGB model with constant SFR of $1 \ \Msun \ \mathrm{yr^{-1}}$, with three different two-component IMFs. The IMF exponent $\alpha$ is set to $\alpha = 1.3$ in the mass range $M =  0.1 \textrm{ -- } 0.5 \ \Msun$ for all of the three IMFs, while three different values $\alpha = 1.3, 2.3 \text{ and } 3.3$ are used in the mass range $M =  0.5 \textrm{ -- } 100 \ \Msun$. The case of such an IMF with exponents $\alpha = (1.3,2.3)$ represents a standard two-component Kroupa IMF \citep{kroupa_variation_2001}. We also include one example of a so-called `paunchy' IMF, given by a three-component power-law with slopes 1, 1.7 and 2.6 in the mass ranges $M =  0.1 \textrm{ -- } 0.5 \ \Msun$, $M =  0.5 \textrm{ -- } 4 \ \Msun$, and $M =  4 \textrm{ -- } 100 \ \Msun$, respectively \citep{fardal_evolutionary_2007}. This kind of IMF leads to a larger midsection of intermediate mass stars. The effect on the Balmer break strength of the different IMF cases are presented in fig~\ref{fig:Bbreak_IMF_PadAgb}. As fig~\ref{fig:Bbreak_IMF_PadAgb} suggests, the Balmer break only becomes significantly stronger in the case where the upper IMF is bottom-heavy ($\alpha=3.3$). Furthermore, the `paunchy' IMF case gives a result that is very similar to the result obtained using a standard Kroupa IMF ($\alpha=2.3$).

\subsection{Comparison to MACS1149-JD1}
\label{sec:JD1}
\begin{figure*}
    \begin{center}
    \includegraphics[width=0.90\textwidth]{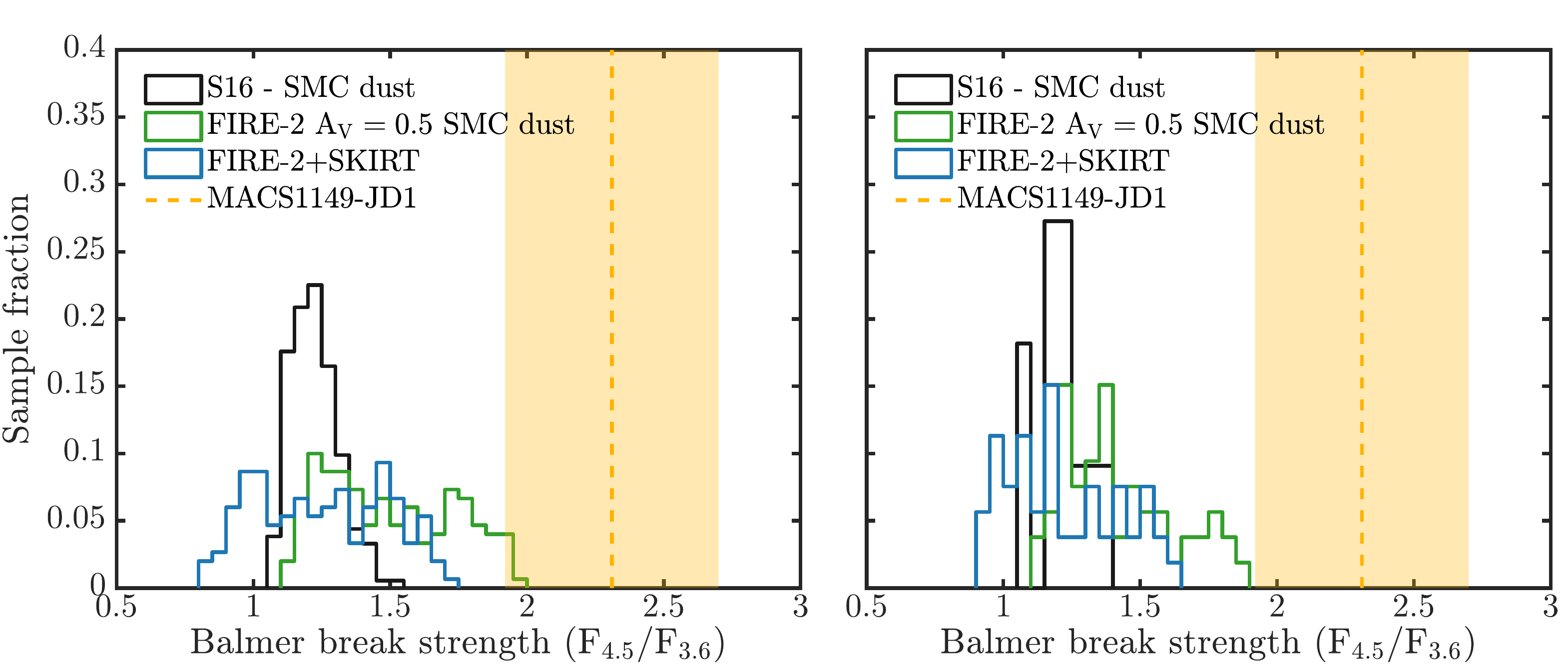}
    \end{center}
    \caption{Balmer break strengths as measured by the \textit{Spitzer}/IRAC $4.5/3.6$-$\mu \mathrm{m}$ flux ratio for simulated galaxies (with $z=9.1$) with $M_{\star} \geq 10^8$ (left) and $M_{\star} \geq 5\times 10^8$ (right) compared to the observed Balmer break in JD1. The yellow dashed line and area show the Balmer break strength and error for JD1 calculated using data from \citet{zheng_young_2017}. The green distribution shows the FIRE-2 galaxies with the Yggdrasil Padova-AGB SSP grid and SMC dust reddening with a fixed extinction of $0.5$ mag in the V-band. The blue distribution shows the FIRE-2 galaxies after post-processing with \textsc{skirt} and the black distributions shows the S16 galaxies with SMC dust extinction.}
    \label{fig:Spitzer_JD1}
\end{figure*}

Fig~\ref{fig:Spitzer_JD1} shows the distribution of Balmer break strengths as measured by \textit{Spitzer}/IRAC for dusty simulated galaxies with masses $M_{\star}>10^8 \ \Msun$ (left panel) and  $M_{\star}>5\times10^8 \ \Msun$ (right panel). Also shown in the figure is the observed Balmer break of JD1, which has a suggested stellar mass of $M_{\star}=1.08^{+0.53}_{-0.18}\times 10^9 \ \Msun$ for a magnification factor of $\mu=10$ \citep{hashimoto_onset_2018}. While the FIRE-2 simulation does produce galaxies that exhibit significantly stronger Balmer breaks than the S16 simulation, due to the more bursty nature of these simulations, none of the simulated galaxies exhibit Balmer breaks as strong as the one seen in JD1. Even in the $A_{\mathrm{V}}=0.5$ SMC dust case, galaxies with Balmer breaks as strong as the one observed in JD1 would be rare. By allowing for a significantly larger magnification than the one suggested for JD1 ($\mu >> 10$), we can include galaxies with lower masses in the comparison, which also leads to a larger sample set. However, even in the case that we include simulated galaxies with masses down to $M_{\star} = 10^8 \ \Msun$, our results are not significantly affected (see left panel in fig~\ref{fig:Spitzer_JD1}).  

As discussed in section~\ref{sec:dust_n_SED-fesc}, increasing the escape fraction may lead to larger Balmer breaks. However, this effect is significantly weaker if the Balmer break is measured through \textit{Spitzer}/IRAC 3.6 and 4.5-$\mathrm{\mu m}$ channel fluxes. Setting the escape fraction to unity leads to an average increase in the measured $\Delta F_{\mathrm{4.5}}/F_{\mathrm{3.6}}$ of $\approx 0.07$ and $\approx 0.05$ for the S16 and FIRE-2 galaxies with the Yggdrasil SSP grid and $A_{\mathrm{V}}=0.5$ SMC dust respectively. The largest increase in individual galaxies is $\Delta F_{\mathrm{4.5}}/F_{\mathrm{3.6}}\approx 0.10$ and $\Delta F_{\mathrm{4.5}}/F_{\mathrm{3.6}}\approx0.15$ for the same two cases with unity escape fraction. In the case that dust-effects are ignored, the largest individual and average increase in the Balmer break for the FIRE-2 galaxies is $\Delta F_{\mathrm{4.5}}/F_{\mathrm{3.6}}\approx 0.14$ and $\Delta F_{\mathrm{4.5}}/F_{\mathrm{3.6}}\approx 0.05$, respectively. We do not see the same clear trend in increasing Balmer Break ($F_{\mathrm{4.5}}/F_{\mathrm{3.6}}$) with initial Balmer break strength as was seen in the case discussed in section~\ref{sec:dust_n_SED-fesc}. If we allow for a dust-to-metal ratio of 0.8 in the \textsc{skirt} post-processing of the FIRE-2 galaxies, we find an insignificant change in the average Balmer break strength as observed with \textit{Spitzer}/IRAC ($\langle\Delta F_{\mathrm{4.5}}/F_{\mathrm{3.6}}\rangle\approx 0.02$). With the largest difference in an individual galaxy of $\Delta F_{\mathrm{4.5}}/F_{\mathrm{3.6}}\approx0.12$. 

\section{Summary and Discussion}
\label{sec:Discussion}
In the following two sections, we discuss our results first from the perspective of future observations with the \textit{JWST} and the Balmer break as probed by $B_{\mathrm{4200/3500}}$ (section~\ref{sec:Disc_BBJWST}). We then go on to discuss our results with respect to JD1, other objects and Balmer breaks as probed by \textit{Spitzer}/IRAC in section~\ref{sec:Disc_JD1}.
\subsection{The Balmer break strength distributions of the simulated galaxies}
\label{sec:Disc_BBJWST}
In the previous sections, we have presented Balmer break distributions as seen in synthetic spectra of simulated galaxies at high redshifts from several different independent simulation suites. By changing parameters such as the dust recipe, SED modelling, IMF assumptions and escape fraction, we are able to get a handle on how these affect the strengths of Balmer breaks exhibited by galaxies in the early universe.

We argue that the upcoming \textit{JWST} should be able to constrain the Balmer breaks of a large number of objects by measuring the continuum flux at 3500 Å and 4200 Å using spectroscopy. In order to get a handle on the feasibility of such observations, we estimate the exposure time required to achieve a signal-to-noise ratio of $5$ in the continuum at 3500 Å and 4200 Å using the \textit{JWST} exposure time calculator\footnote{\url{https://jwst.etc.stsci.edu}} \citep{pontoppidan_pandeia:_2016}. In this procedure, we use SEDs of ten $z=9$ galaxies from the S16 simulation with average UV magnitudes and re-scale these to $\mathrm{m_{AB}(1500\text{Å})}=27 \text{ mag}$. We find that for the \textit{JWST}/NIRSpec low resolution setting ($R\sim 100$) we should achieve a signal-to-noise ratio per spectral bin of $5$ in $\approx 30$ hours for a galaxy with $\mathrm{m_{AB}(1500\text{Å})}=27 \text{ mag}$. However, even in this low-resolution setting, there are several spectral bins ($\sim 5$) that should be free of absorption/emission lines. By combining these we should be able achieve the same signal to noise within $\approx 10$ hours. The exact exposure time required of course depends on the shape of the continuum and size of the Balmer break. In principle, it should be straightforward to compare observed distributions of Balmer break strengths to the ones presented in this study in order to get a handle on how well the simulations and models match observations and thereby get a handle on how star formation proceeded in the early universe. We show that in cases where the escape fraction of ionizing photons is extreme, this can have a significant effect on the Balmer break strength in the simulated galaxies. This effect can be attributed to the weaker nebular continuum around 3500 Å at higher escape fractions. This effect is larger in galaxies that exhibit the weakest Balmer breaks, since these contain more young stars and thus exhibit more nebular continuum emission on the blue side of the Balmer break. This means that one can, in principle, treat the Balmer break as a diagnostic of age, and use the Balmer break strength distributions to look for the youngest objects \citep[e.g.][]{raiter_predicted_2010, inoue_rest-frame_2011}. 
As shown in section~\ref{sec:dust_n_SED-Dust}, while the general effect of dust is to increase the strength of the Balmer breaks of the simulated galaxies, the effect on the overall distribution of Balmer break strengths is small unless an extreme or very specific scenario is employed. While this does not rule out a case in which dust leads to extreme Balmer breaks \citep{katz_probing_2019}, we find that such cases are uncommon in our simulations. In the case of dust reddening of the S16 and FIRE-2 galaxies where the Yggdrasil SSP grid were used, we only apply the SMC extinction curve. One could, in principle, use different dust reddening curves such as the \citet{calzetti_dust_2000} attenuation law for comparison. However, a flatter extinction like the Calzetti law is unlikely to produce stronger Balmer breaks; thus, the SMC law allows us to discuss the case in which dust is most likely to produce a strong Balmer break. We have also utilized a Yggdrasil SSP grid which has been extended to lower metallicity using models from \citet{raiter_predicted_2010} at $Z=10^{-7}\textrm{ -- }10^{-5}$. We find, however, that the difference in the Balmer breaks produced by extending the grid to lower metallicity is insignificant.

\subsection{Simulated galaxies and MACS1149-JD1}
\label{sec:Disc_JD1}
It has been suggested that the prominent Balmer break observed in JD1 is a result of the galaxies' peculiar SFH. In this scenario, the bulk of the stars in the galaxy are produced in an early episode of star formation, which is followed by a quite long period of low star formation activity before a second burst. Of course, there are other scenarios that could produce strong Balmer breaks. For example, less extreme variations in the star formation activity coupled with dust-reddening could lead to stronger Balmer breaks depending on the exact properties and distribution of the dust \citep{katz_probing_2019}. In principle, one could also imagine scenarios where the shape of the IMF leads to significantly stronger Balmer breaks. Here, we have used several simulation suites and varied assumptions regarding the dust properties, shape of the IMF and the escape fraction of ionizing photons in order to get a handle on which effects may lead to strong Balmer breaks. The results from our simulations suggest that galaxies such as JD1 should be rare in the early universe, as our simulations produce galaxies that exhibit significantly weaker Balmer breaks than the one observed in JD1. We find that while factors such as the escape fraction of LyC photons and stronger dust extinction may lead to larger Balmer breaks, we find no likely scenarios that produce simulated SEDs with Balmer breaks as strong as the one observed in JD1. We also find that allowing for stronger magnification than the one observed in JD1 does not significantly change our results. Other factors, such as adopting a flatter extinction curve than the SMC curve used here, or using SSP models that include stellar binarity may, furthermore, lead to slightly smaller Balmer breaks. We find that a dust-screen model does not lead to very strong Balmer breaks unless some extreme scenario is employed. In the case of an galaxy-wide SMC extinction with $A_{\mathrm{V}}=0.5$, none of the simulated galaxies exhibit Balmer breaks as strong as the one observed in JD1. In regards to the IMF, we find that the Balmer break only becomes significantly stronger in cases in which a bottom-heavy IMF is employed.

\citet{katz_probing_2019} show that they are able to reproduce the observed Balmer break of JD1 without the need of extreme variations in the SFR. While their simulated galaxies show variations in the star formation activity, these exhibit strong Balmer breaks mainly as an effect of strong differential obscuration, where the light of young stars is largely extinguished by dust. Indeed, we do expect dust to be concentrated in regions of higher density where star formation is occurring, and where young stars are likely to be found \citep[see e.g.][]{charlot_simple_2000}. This effect is also observed in the FIRE-2 galaxies that have been post-processed with \textsc{skirt}. Even so, we find no galaxies in which this effect is strong enough to reproduce the Balmer break observed in JD1. Contrary to the dust prescription used in \citep{katz_probing_2019}, the \textsc{skirt} post-processing also considers scattering by dust. In principle, light scattered into the line-of-sight may contribute significantly to the post-extinction flux in the simulated galaxies \citep{ma_dust_2019}. Thus, the effect of differential obscuration may become significantly smaller in cases where dust scattering is considered. In addition to finding an analogue to JD1, \citet{katz_probing_2019} show that they find 3 objects that exhibit strong Balmer breaks in their simulation volume. It is, however not entirely clear how likely we are to find such galaxies in their simulations when considering a larger sample of objects (especially when considering possible viewing-angle effects). Furthermore, \citet{katz_probing_2019} point out that strong Balmer breaks are rare in their simulations if one considers dust-free galaxies, which is consistent with our findings. In difference to the \citet{katz_probing_2019} study, we find that galaxies with Balmer breaks as strong as the one observed in JD1 are rare in the simulations at $z\sim 9$, even when considering differential obscuration. This means that the main difference in our results stem from the different treatment of dust effects. Note, that while we find that JD1 would be a rare type of object in the simulations presented here, this does not rule out that one could find an analogue of JD1 given larger simulation volumes/samples.

Is it possible that JD1 is an outlier in the intrinsic Balmer break distribution due to selection effects? JD1 was targeted with ALMA due to the photometric redshift prediction, which places the [OIII] 88-$\mathrm{\mu m}$ emission at a fitting wavelength for ALMA. In principle, it could be the case that the photometric redshift determination of JD1 hinges on a strong Balmer break, and that we are in this way biased to find galaxies such as JD1 at the estimated redshift.  We performed tests using the \textsc{eazy} photometric redshift code \citep{brammer_eazy:_2008} for measurements from \citet{zheng_young_2017} in order to understand if the Balmer break strongly affects the redshift determination. We ran \textsc{eazy} using the built-in prior (and with no prior), with several of the built-in libraries, while altering the \textit{Spitzer}/IRAC 4.5 $\mu m $ measurement, or removing it completely. We found no significant difference in the obtained best photometric redshift or posterior distributions. Thus, we see no reason to think we are biased to find objects such as JD1 in the high-redshift universe. This suggests that JD1 is either a common type of object, or it has been drawn from the tail of the galaxy distribution basically due to `bad luck'. Given the large error-bar on the \textit{Spitzer}/IRAC measurements of the Balmer break in JD1, and the fact that JD1 represents only one system, conducting similar observations for a wider class of high-redshift objects should be of high priority. As mentioned in the introduction (section \ref{sec:intro}), while there is a lack of spectroscopically confirmed galaxies at these highest redshifts, there are several galaxies which have photometric redshifts above 9.1 and that exhibit similar \textit{Spitzer}/IRAC colors as JD1. Looking at the combined sample  \citet{bouwens_newly_2019} and \citet{oesch_most_2014} we find 11 objects with $z_{phot}\geq 9.1$ \citep[including GN-z11, which has a measured spectroscopic redshift;][]{oesch_remarkably_2016}. Out of these, there are a few objects that exhibit red \textit{Spitzer}/IRAC colors which are comparable to JD1 ($F_{4.5}/F_{3.6}\gtrsim 2$). It is important to note that the lack of a spectroscopic redshift increases the uncertainty regarding the nature of the red color in these channels, since contribution to from emission lines such as [OIII]~$\lambda$5007 to the 4.5-$\mathrm{\mu m}$ channel cannot be ruled out. However, if future observations are able to confirm that the red color does in fact arise due to strong Balmer breaks, this would indeed mean that objects like JD1 are much more common than predicted by the simulations used here. If this turns out to be the case, this may be an indication that simulations and models are missing some key physical ingredient that is required to reproduce observations at the highest redshifts. For example, stronger feedback effects in the simulations should be more likely to produce larger variations in the SFR and ultimately Balmer break strengths, and thus could be more likely to produce Balmer breaks as strong as the one seen in JD1. Ultimately, the \textit{JWST} should be able to shed some light on the nature of JD1 along with the galaxy population at these very high redshifts, helping us determine if JD1 represents a common type of object in the early universe, and allow us to calibrate and/or verify our models.

\section*{Acknowledgements}
EZ and AV acknowledge funding from the Swedish National Space Board. AKI and TH acknowledge the financial support from NAOJ ALMA Scientific Research Grant number 2016-01 A. AKI also acknowledges the support from JSPS KAKENHI 17H01114. TH also acknowledges the Grant-inAid for Scientific Research 19J01620. Numerical simulations of S16 have been performed with Cray XC30, XC50 in CfCA at NAOJ and with Cray XC40 at YITP in Kyoto University. DC is a DAWN fellow. This research has made use of NASA’s Astrophysics Data System.




\bibliographystyle{mnras}
\bibliography{BalmerBreak} 



\appendix

\section{Additional figures}
\begin{figure}
\begin{center}
\includegraphics[width=0.48\textwidth]{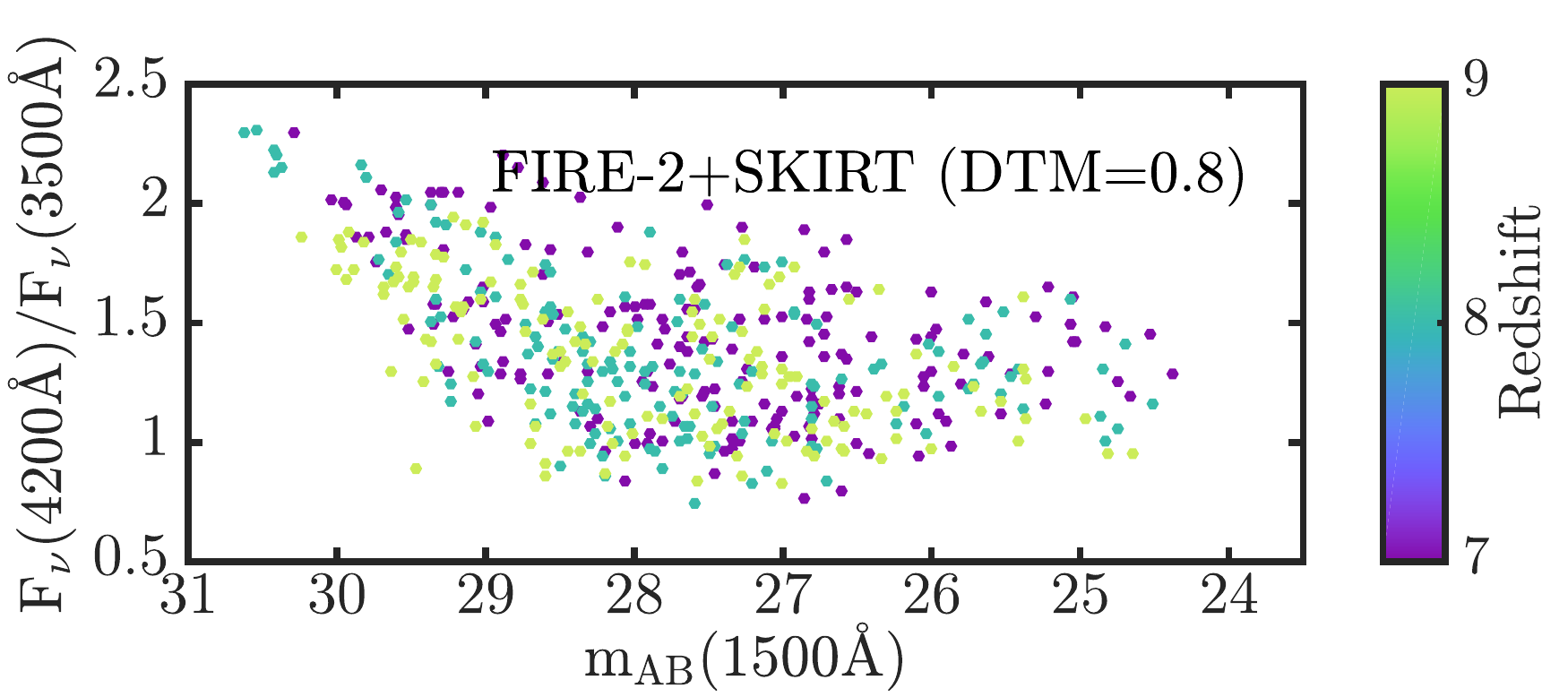}
\end{center}
\caption{UV magnitude ($\mathrm{m_{AB}(1500\text{Å})}$) vs Balmer break strength ($ F_{\nu}(4200 \text{Å})/F_{\nu}(3500 \text{Å}) $, $ B_{\mathrm{4200/3500}}$ in the text) vs redshift (color axis) for the FIRE-2 simulations that have undergone post-processing in \textsc{skirt} assuming a dust-to-metal ratio of 0.8}
\label{fig:UVmag_FIRE2_dtm08}
\end{figure}

Fig~\ref{fig:UVmag_FIRE2_dtm08} shows $ B_{\mathrm{4200/3500}}$ as a function of observed UV magnitude for the FIRE-2 galaxies that have been post-processed using \textsc{skirt} assuming a dust-to-metal ratio of 0.8 (see section~\ref{sec:Synthetic_SEDs}). 

\label{sec:appendix}


\bsp	
\label{lastpage}
\end{document}